\documentclass[preprint,12pt,number]{elsarticle}
\journal{Journal of Non-Newtonian Fluid Mechanics}

\usepackage[T1]{fontenc}
\usepackage[utf8]{inputenc}
\usepackage[english]{babel}
\usepackage{amsmath,amssymb,bm}
\usepackage{graphicx}
\usepackage{booktabs}
\usepackage{siunitx}
\usepackage{xcolor}

\usepackage[colorlinks=true,linkcolor=blue!50!black,citecolor=blue!50!black,urlcolor=blue!50!black]{hyperref}

\newcommand{\bv}{\bm{u}}
\newcommand{\D}{\bm{D}}
\newcommand{\sq}{\sigma_q}
\newcommand{\st}{\sigma_{\mathrm{tot}}}
\newcommand{\Pe}{\mathit{Pe}}
\newcommand{\Rey}{\mathit{Re}}
\newcommand{\St}{\mathit{St}}
\newcommand{\Cu}{\mathit{Cu}}
\newcommand{\dd}{\,\mathrm{d}}

\begin{document}

\begin{frontmatter}

\title{Shear thinning confines the wall forcing: laminar chaotic mixing of Carreau fluids in a rotating arc-wall mixer}

\author{Teodor Burghelea}
\ead{Teodor.Burghelea@univ-nantes.fr}
\address{Nantes Universit\'e, CNRS, Laboratoire de Thermique et \'Energie de Nantes, LTeN, UMR 6607, F-44000 Nantes, France}

\begin{abstract}
With two-dimensional simulations of Carreau fluids of power-law index $N$ from 1 to 0.4, I study how shear thinning modifies laminar chaotic mixing in an open channel stirred by three arc-shaped wall segments rotating back and forth periodically, the upper one in anti-phase with the two lower ones. The outlet mixing efficiency falls monotonically, by 14\,\% from $N=1$ to 0.4, mostly between $N=0.85$ and 0.6. The bulk flow and its chaotic saddle are nearly unaffected. Shear thinning acts instead through a thin layer of reduced viscosity around the moving arcs: the geometric penetration length of the wall forcing shortens as the 0.3 power of the time-mean wall viscosity, within 4\,\% over a fourfold range of forcing. This confinement weakens the stretching, and the penetration length predicts the mixing efficiency ($r^2=0.97$). The optimal Strouhal number stays close to 0.7 for all fluids, but shear thinning narrows the range of efficient periods. A lamellar model indicates that the relative penalty of shear thinning grows with the P\'eclet number. As the injected power scales with the wall viscosity, the mixing efficiency per unit energy rises 2.6-fold from $N=1$ to 0.4, and at equal energy per unit volume, with a faster rotation, the fluid with $N=0.4$ mixes as well as the Newtonian one, and better at larger energies. In wall-driven mixers, the reach of the forcing, not the bulk viscosity, is the quantity to control.
\end{abstract}


\end{frontmatter}

\section{Introduction}\label{sec:intro}

Many materials of industrial relevance --- polymer melts and solutions, food pastes, cosmetic and pharmaceutical formulations, slurries --- must be mixed while they are highly viscous, where turbulence is either unattainable within reasonable energy budgets or undesirable, because its large stresses can damage a fragile microstructure. Mixing must then rely on laminar flows that stretch and fold fluid elements repeatedly until striations are thin enough for diffusion to complete the homogenisation. Since the work of Aref \cite{aref1984} it is known that even simple time-periodic laminar flows can generate chaotic trajectories, a regime known as chaotic advection \cite{ottino1989,aref2002,aref2017}, whose kinematic foundations --- exponential stretching, folding, and the competition between stretching and diffusion --- are well established \cite{diffusiontime,Stretchingandfolding,wiggins2004}.

Most industrial operations are continuous, and the relevant setting is that of open flows, in which fluid crosses a region of chaotic stirring in a finite time. Mixing is then controlled by transient chaos: fluid entering the mixer spends a distributed time near a chaotic saddle and is stretched by an amount that depends on this residence time \cite{tel2005,dmfg2012,lt2011}. Residence-time distributions of in-line chaotic mixers have been characterised \cite{PhysRevE.106.015107}, and no-slip walls are known to slow down the decay of the scalar variance by trapping poorly stretched fluid \cite{gouillart2007,gouillart2008}.

The rotating arc-wall mixer \cite{elomari2009,PhysRevFluids.6.024502} belongs to this class. A plane channel is equipped with three arc-shaped segments of its walls which rotate periodically about their centres, so that the through-flow is stirred by moving boundaries rather than by internal rods or baffles. Simulations \cite{PhysRevFluids.6.024502} and experiments with Newtonian fluids \cite{younes2021} have shown that it mixes efficiently for suitable periods and strengths of the rotation, controlled by a Strouhal number and by the ratio of the wall speed to the mean flow, with an optimal forcing frequency \cite{younes2021}. The concept has been extended to viscoplastic materials \cite{youne2020,kiwan2025,moguen2023} and, experimentally, to shear-thinning fluids \cite{kiwan2025}. It carries over to an open channel the alternately rotating walls of closed thermal mixers \cite{elomari2010a,leguer2012}; alternating between two flows whose streamlines cross is the elementary mechanism of such devices, of which linked twist maps are the minimal model \cite{wiggins2004}.

How shear thinning modifies chaotic mixing has received comparatively little attention. In the flow between eccentric cylinders rotated alternately, Niederkorn and Ottino \cite{ottino1} showed that departures of a few per cent from the Newtonian velocity field, caused by a viscosity drop next to the moving cylinder, change chaotic advection markedly: hyperbolic periodic points can become elliptic, creating regular islands, and the stretching slows down, although a suitable choice of the operating conditions restores good mixing. They also noted that, whereas a Newtonian Stokes flow depends only on the displacement of the walls per period \cite{swanson1990}, with a shear-thinning fluid the wall speed matters too, because it sets the viscosity. Shear thinning and viscoelasticity were compared in a model two-dimensional mixer \cite{connelly2004}, and thermal chaotic mixing of power-law fluids was studied with alternately rotating walls \cite{elomari2010b}. In these closed flows a modest change of the velocity field acts mainly through bifurcations of the flow topology.

An open wall-driven mixer differs in two respects: the fluid is stirred only during its transit, and the momentum that drives the stirring is injected through short moving wall segments, where the shear rates, and therefore the viscosity reduction, are largest. Shear thinning may thus act directly on the transmission of the forcing to the bulk. Experiments in the rotating arc-wall mixer with two carboxymethylcellulose solutions, of power-law indices 0.76 and 0.57 \cite{kiwan2025}, showed that they mix as well as a Newtonian fluid under slow forcing but less well above an intermediate Strouhal number, with thicker poorly mixed wall layers and lower finite-time Lyapunov exponents; the degradation was attributed to a viscous slip near the rotating walls. In experiments, however, the power-law index cannot be changed without changing the other rheological parameters, and the mechanism has not been quantified.

In this paper I simulate the mixing of Carreau fluids in the rotating arc-wall mixer for thirteen power-law indices between $N=1$ and 0.4, all other parameters being fixed, so that $N$ is the only control parameter. The arc rotation is imposed as a moving-wall condition on embedded boundaries, so that the near-wall layer, where the viscosity varies most, is resolved directly. I relate the outlet mixing to the Eulerian flow structure, to the transmission of the wall forcing into the bulk and to the Lagrangian residence times and stretching, and show that shear thinning leaves the bulk stirring flow essentially unchanged but confines the wall forcing within a thinned near-wall layer, following an explicit scaling with the wall viscosity which predicts the mixing efficiency at a given forcing and its energetic cost. Sec.~\ref{sec:methods} presents the problem, the numerical method and the diagnostics; Secs.~\ref{sec:mixing}--\ref{sec:lagrangian} describe the effect of shear thinning on mixing, on the flow and on the Lagrangian transport; Secs.~\ref{sec:discussion} and \ref{sec:conclusions} discuss and summarise the results.

\section{Problem setting and methods}\label{sec:methods}

\subsection{Geometry and forcing}\label{subsec:geometry}

\begin{figure}[htbp]
\centering
\includegraphics[width=\textwidth]{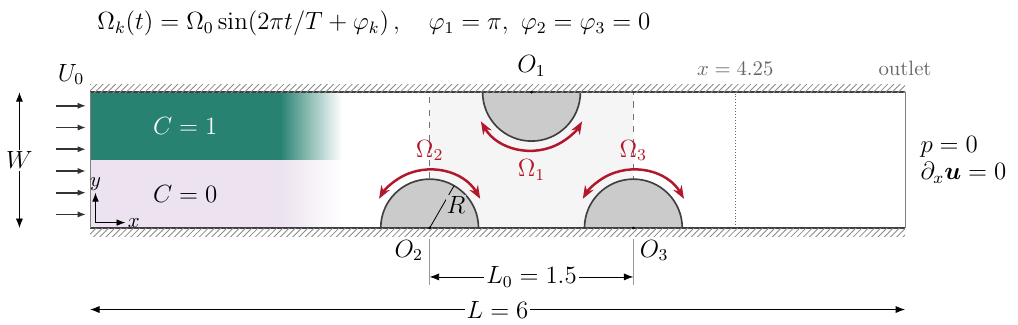}
\caption{The rotating arc-wall mixer (to scale). Two streams enter with a uniform velocity $U_0$, carrying the scalar at $C=1$ (upper half) and $C=0$ (lower half). The three arcs of radius $R$ rotate periodically about $O_{1,2,3}$, the bottom arcs in phase and the top arc in anti-phase with them; the shaded region of length $L_0$ is the active mixing zone. The dotted lines mark the stations $x=4.25$ and $x=5.5$ (outlet) where the mixing is evaluated.}
\label{fig:geometry}
\end{figure}

The mixer (Fig.~\ref{fig:geometry}) is a plane channel of width $W=1$ and length $L=6$, $x\in[-0.5,5.5]$, $y\in[-W/2,W/2]$. Three circular arcs of radius $R=0.36$, centred on the walls at $O_1=(2.75,\,0.5)$ (top) and $O_2=(2,\,-0.5)$, $O_3=(3.5,\,-0.5)$ (bottom), protrude into the channel by $R$; $L_0=1.5$ is the length of the active mixing zone.

Each arc rotates rigidly about its centre with the angular velocity
\begin{equation}\label{eq:forcing}
\Omega_j(t)=\Omega_0\,\sin\!\left(\omega t+\varphi_j\right),\quad \omega=\frac{2\pi}{T},\qquad \varphi_2=\varphi_3=0,\quad \varphi_1=\pi ,
\end{equation}
with $\Omega_0=150$ and $T=0.5$, so that the maximum wall speed is $\Omega_0 R=54$. This is the protocol of the experiments \cite{younes2021,kiwan2025}, selected for repeatedly creating and destroying elliptic regions \cite{youne2020}. The other walls are at rest. The fluids enter at $x=-0.5$ with a uniform velocity $U_0=2$, the upper half carrying the scalar at $C=1$ and the lower half at $C=0$; a zero-pressure, zero-normal-gradient condition is applied at the outlet $x=5.5$. At $t=0$ the fluid is at rest, with $C=1$ for $y>0$ and $C=0$ for $y<0$.

\subsection{Governing equations and rheology}\label{subsec:equations}

The flow is incompressible and two-dimensional, and the concentration $C$ of a passive scalar obeys an advection--diffusion equation:
\begin{align}
\nabla\cdot\bv &= 0, \label{eq:continuity}\\
\rho\left(\partial_t\bv+\bv\cdot\nabla\bv\right) &= -\nabla p+\nabla\cdot\left[2\,\eta(\dot\gamma)\,\D\right], \label{eq:momentum}\\
\partial_t C+\bv\cdot\nabla C &= \mathcal{D}\,\nabla^2 C, \label{eq:scalar}
\end{align}
where $\bv=(u,v)$ is the velocity, $p$ the pressure, $\D=\tfrac12\left[\nabla\bv+(\nabla\bv)^{T}\right]$ the rate-of-strain tensor, $\rho=1$ and $\mathcal{D}=10^{-7}$. No-slip conditions hold on all walls, and the scalar satisfies a zero-flux condition. The fluids follow the Carreau model \cite{carreau1972,Bird77}
\begin{equation}\label{eq:carreau}
\eta(\dot\gamma)=\eta_\infty+(\eta_0-\eta_\infty)\left[1+(\lambda\dot\gamma)^2\right]^{\frac{N-1}{2}},
\end{equation}
with $\eta_0=1$, $\eta_\infty=0.1$, $\lambda=0.01$ and $N\in\{0.4,0.45,\dots,0.95,1\}$. The shear rate in Eq.~\eqref{eq:carreau} is $\dot\gamma=2\sqrt{\D:\D}$, i.e.\ $\sqrt2$ times the conventional $\sqrt{2\,\D:\D}$ \cite{Bird77}; equivalently, the results correspond to the conventional shear rate with a relaxation time $\sqrt2\,\lambda\simeq0.014$.

\subsection{Dimensionless groups}\label{subsec:groups}

With $W$, $U_0$ and $\eta_0$ as reference scales, the dimensionless groups are those of Table~\ref{tab:groups}. The mean flow is slow ($\Rey=2$) whereas the forcing is strong: the arc surfaces move 27 times faster than the mean flow. $\St$ and $\Lambda$ are defined as in the experiments \cite{kiwan2025}, where $\Lambda$ is denoted $K$. The Carreau number $\Cu=\lambda\Omega_0$ indicates that shear thinning is significant at the shear rates imposed by the arcs.

\begin{table}[htbp]
\centering
\caption{Dimensionless groups (values common to all fluids).}
\label{tab:groups}
\begin{tabular}{@{}llr@{}}
\toprule
Group & Definition & Value\\
\midrule
Reynolds number (mean flow) & $\Rey=\rho U_0 W/\eta_0$ & 2\\
Reynolds number (wall forcing) & $\Rey_w=\rho\,\Omega_0 R\,W/\eta_0$ & 54\\
Forcing ratio & $\Lambda=\Omega_0 R/U_0$ & 27\\
Strouhal number & $\St=W/(U_0T)$ & 1\\
P\'eclet number & $\Pe=U_0W/\mathcal{D}$ & $2\times10^{7}$\\
Carreau number & $\Cu=\lambda\,\Omega_0$ & 1.5\\
Viscosity ratio & $\eta_\infty/\eta_0$ & 0.1\\
Arc radius & $R/W$ & 0.36\\
Power-law index & $N$ & 0.4--1\\
\bottomrule
\end{tabular}
\end{table}

\subsection{Numerical method}\label{subsec:numerics}

Equations \eqref{eq:continuity}--\eqref{eq:scalar} are solved with Basilisk \cite{basilisk}, which descends from Gerris \cite{gerris1,popinet2009}, on an adaptive quadtree grid with a second-order time-split projection method, Bell--Colella--Glaz advection \cite{bcg1989} and implicit viscous and diffusive terms. The channel walls and the arcs are embedded boundaries cutting the Cartesian cells \cite{johansen1998}, on which the velocity of the arc surfaces is imposed as a Dirichlet condition. The grid is adapted at every step with a wavelet criterion on the solid fraction, the velocity and the scalar \cite{popinet2015}; the finest cells have size $h=6/2^{10}\simeq5.9\times10^{-3}$ (171 cells across the channel), and the grid is uniformly fine across the thinned layer next to the arcs. The time step (CFL 0.8, $\Delta t\le10^{-3}$) varies between about $6\times10^{-5}$ and $7\times10^{-4}$. Each simulation covers $0\le t\le10$ (twenty periods), with fields stored every $0.01\,T$.

The velocity becomes time-periodic within three periods (periodicity defect below $10^{-2}$, 0.1--0.3\,\% over the last five periods) and the scalar after the initial interface has been swept out, at $t\simeq4.5$. All statistics are computed over the window $5\le t\le10$; shifting it by five periods changes $M$ by about $2\times10^{-3}$, well below the differences between the fluids. The flow rate is conserved within 0.3\,\%, and the over- and undershoots of the scalar at the sharpest fronts affect 0.2--0.4\,\% of the outlet section; $C$ is clipped to $[0,1]$, which changes the outlet statistics by less than 0.4\,\%. The adapted grid, the establishment of the periodic regime and these checks are documented in the supplementary material (Secs.~S1--S2).

\paragraph{Grid convergence}\label{subsec:grid} The two extreme fluids, $N=1$ and 0.4, were also simulated with maximum levels 9 and 11. The flow is converged at level 10: from level 10 to 11 the penetration length (Sec.~\ref{subsec:diagnostics}) changes by 0.05\,\% for $N=1$ and 0.6\,\% for $N=0.4$, the wall viscosity by 3\,\%, and the dissipation by 5--7\,\%, mostly at the feet of the arcs where the stress is singular. The scalar is not converged, as expected at $\Pe=2\times10^7$: striations are homogenised once they reach the grid scale, and the outlet mixing efficiency $M$ decreases with refinement (0.700, 0.654 and 0.622 for $N=1$, and 0.602, 0.562 and 0.515 for $N=0.4$, at levels 9, 10 and 11). The effect of shear thinning depends little on the resolution: $M(0.4)/M(1)=0.860$ at levels 9 and 10 and 0.829 at level 11, so that the penalty reported below (14\,\%) is, if anything, a slight underestimate. In a companion study of the Newtonian fluid \cite{burghelea_nwt}, runs at resolved P\'eclet numbers and a lamellar model show that numerical diffusion at level 10 acts as a molecular diffusivity with an effective P\'eclet number $\Pe_g\approx2\times10^4$. The conclusions below therefore rest on the variation of $M$ with $N$ at fixed resolution and on Lagrangian measures computed from the converged velocity field; the consequences of a larger P\'eclet number are discussed in Sec.~\ref{sec:lagrangian}.

\subsection{Diagnostics}\label{subsec:diagnostics}

\paragraph{Mixing indicator} The mixing is measured at the outlet $x_0=5.5$ by the flux-weighted standard deviation of the concentration taken over the channel width \emph{and} over the periodic regime,
\begin{equation}\label{eq:sigmatot}
\st=\left[\frac{\iint u\,(C-\bar C)^2\dd y\dd t}{\iint u\dd y\dd t}\right]^{1/2},\qquad
\bar C=\frac{\iint u\,C\dd y\dd t}{\iint u\dd y\dd t},
\end{equation}
i.e.\ the deviation of the fluid collected in a cup at the outlet over $5\le t\le10$. Perfect segregation gives $\st=1/2$ and perfect mixing $\st=0$; the mixing efficiency is $M=1-2\st$. In the absence of diffusion, $\iint uC^2\dd y\dd t$ is the same on every section of a time-periodic flow, so that $\st$ can only decrease downstream, as it does in all the simulations. With the instantaneous cup-mixing deviation on a section,
\begin{equation}\label{eq:sigmaq}
\sq(x_0,t)=\left[\frac{\int u\,(C-\bar C_q)^2\dd y}{\int u\dd y}\right]^{1/2},\qquad
\bar C_q(x_0,t)=\frac{\int u\,C\dd y}{\int u\dd y},
\end{equation}
and $q(t)=\int u\dd y$, the total variance splits exactly into a spatial and a temporal part,
\begin{equation}\label{eq:split}
\st^2=\underbrace{\frac{\int q\,\sq^2\dd t}{\int q\dd t}}_{\text{spatial}}+\underbrace{\frac{\int q\,(\bar C_q-\bar C)^2\dd t}{\int q\dd t}}_{\text{temporal}},
\end{equation}
the temporal part measuring the alternation of rich and lean pulses released by the arcs. Downstream of the arcs the velocity profile converts the temporal part into the spatial one, so that $\sq$ alone increases towards the outlet although the fluid does not unmix (supplementary Sec.~S3). Behind the last arc ($x\lesssim4.5$) 11--22\,\% of the flux is reversed, which makes single-section indicators ill-defined there.

\paragraph{Flow type} The local nature of the flow is characterised by
\begin{equation}\label{eq:flowtype}
\xi=\frac{\lVert\D\rVert-\lVert\bm{W}\rVert}{\lVert\D\rVert+\lVert\bm{W}\rVert},
\end{equation}
where $\bm W$ is the vorticity tensor; $\xi=1$, 0 and $-1$ correspond to pure extension, simple shear and rigid rotation \cite{FULLER1981271}.

\paragraph{Penetration of the wall forcing} The amplitude of the oscillating velocity, $u'_{\mathrm{rms}}(\bm x)=\big[\overline{(u-\bar u)^2}+\overline{(v-\bar v)^2}\big]^{1/2}$ (overbar: time average over the periodic regime), is binned as a function of the distance $d$ to the nearest arc surface within the active zone. The penetration length $\ell$ is the e-folding length of an exponential fit of $u'_{\mathrm{rms}}(d)$ over $0.01\le d\le0.1$, i.e.\ between two cells from the wall and the distance at which the neighbouring arcs become felt. Other definitions change $\ell$ by up to a third but not its dependence on $N$ (Sec.~\ref{subsec:penetration}).

\paragraph{Energy budget} The power injected by the arcs, $P_{\mathrm{in}}=\sum_j\int_{\mathrm{arc}\,j}\tau_w\,\Omega_j R\dd S$, and the viscous dissipation $\Phi=\int 2\eta\,\D:\D\dd A$ are averaged over the periodic regime. The wall stress $\tau_w=\eta_w\,\partial u_{\mathrm{rel}}/\partial r|_{r=R}$ is computed from the velocity relative to the rigid rotation of the arc, by a one-sided quadratic through the velocities at $1.5h$ and $3h$ from the surface. The kinetic-energy flux through the boundaries and the pressure work of the mean flow are below 1\,\% of $P_{\mathrm{in}}$.

\subsection{Lagrangian diagnostics}\label{subsec:lagr_methods}

\paragraph{Velocity field and trajectories} The ten periods $5\le t<10$ are phase-averaged into a time-periodic velocity field stored at 100 phases per period on the uniform level-10 grid, the arc interiors being filled with their rigid rotation. Trajectories are integrated with a fourth-order Runge--Kutta scheme ($\Delta t=5\times10^{-4}$; halving it changes the mean stretching by 0.2\,\%), with bilinear interpolation in space and linear in time; tracers that numerically cross a wall are reflected back into the fluid.

\paragraph{Residence times and stretching} Tracers are released on the line $x_{\mathrm{in}}=1$ at 1000 points across the channel and 20 phases, each weighted by the local flux $u(x_{\mathrm{in}},y_0,t_0)$, and followed until they cross $x_{\mathrm{out}}=4.75$, downstream of the recirculation behind the last arc, or for at most twelve periods. The flux-weighted density $E(\tau)$ of the transit time $\tau$ is the residence-time distribution in the sense of Danckwerts \cite{danckwerts1953,nauman2008}, whose mean is fixed by mass conservation, $V/Q=1.57$. The reference is the plane Poiseuille flow of the same mean $\bar\tau$, for which, with $\theta=\tau/\bar\tau$,
\begin{equation}\label{eq:rtd_pois}
F_P(\theta)=\tfrac12\sqrt{1-\tfrac{2}{3\theta}}\,\Big(2+\tfrac{2}{3\theta}\Big),\qquad \theta\ge\tfrac23,
\end{equation}
with an algebraic tail $E_P\sim\theta^{-3}$ due to the no-slip walls \cite{gouillart2007}. Each tracer carries an infinitesimal material line element $\bm{\delta l}$, initially at $45^\circ$ to the axis, with $\mathrm{d}\bm{\delta l}/\mathrm{d}t=(\nabla\bm u)^{\mathsf T}\!\cdot\bm{\delta l}$; its log-stretch on exit,
\begin{equation}\label{eq:logstretch}
s=\ln\frac{|\bm{\delta l}(\tau)|}{|\bm{\delta l}(0)|},
\end{equation}
is the Lagrangian counterpart of the striation thinning: a filament of width $\delta_0$ leaves with width $\delta_0e^{-s}$ \cite{ottino1989}, and $s/\tau$ is the mean stretching rate. Stretching statistics predict the spreading of a tracer better than stroboscopic sections \cite{swanson1990}; I therefore rank the fluids by the distribution of $s$ and use the sections qualitatively.

\paragraph{Finite-time Lyapunov exponents} For the flow map $\bm\phi_{t_0}^{t_0+\Delta}$,
\begin{equation}\label{eq:ftle}
\sigma_{t_0}^{\Delta}(\bm x_0)=\frac{1}{|\Delta|}\ln\sqrt{\lambda_{\max}\big(\bm C\big)},\qquad
\bm C=\big(\nabla_0\bm\phi_{t_0}^{t_0+\Delta}\big)^{\mathsf T}\nabla_0\bm\phi_{t_0}^{t_0+\Delta},
\end{equation}
where $\lambda_{\max}$ is the largest eigenvalue of the Cauchy--Green tensor $\bm C$ \cite{slm2005,haller2015}. Initial positions of spacing $2h$ over $1.5\le x\le4.5$ are advected over $\Delta=T$ and $\pm2T$ from phase 0. Ridges of the forward and backward fields mark the stable and unstable manifolds of the chaotic set, which bound the lobes exchanged between the mixing zone and its surroundings \cite{branicki2010}.

\paragraph{Chaotic saddle} In an open flow almost every particle leaves the mixing zone after a finite time, and the long-lived particles accumulate on a fractal chaotic saddle and its unstable manifold \cite{jung1993,lt2011}. A grid of $8.8\times10^4$ tracers of spacing $\simeq h$ is seeded over $1\le x\le4.75$ at phase 0 and advected, tracers being removed at $x_{\mathrm{out}}$; the survivors at $t=nT$, $n\ge4$, form stroboscopic sections. The survival probability decays as $P_s\sim e^{-\kappa t}$ for a hyperbolic saddle, which defines the escape rate $\kappa$ \cite{kantz1985}; the Lyapunov exponent of the saddle, $\lambda_s$, is the mean growth rate of the log-stretch of the survivors. For an area-preserving map the Kantz--Grassberger relation gives the dimension of the unstable manifold, $D_u=2-\kappa/\lambda_s$, between 1 (fast escape, weak chaos) and 2 (area-filling, strong chaos) \cite{kantz1985,lt2011}.

\section{Effect of shear thinning on mixing}\label{sec:mixing}

\subsection{Scalar patterns}
\begin{figure}[htbp]
\centering
\includegraphics[width=\textwidth]{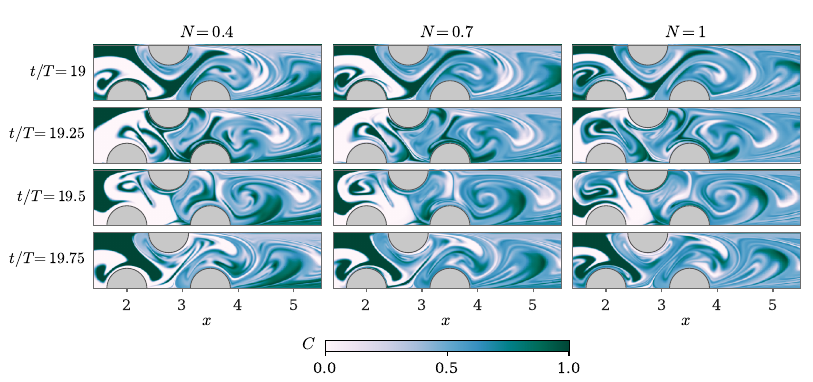}
\caption{Concentration field $C$ downstream of $x=1.4$ at four phases of the last forcing period, for three values of $N$; grey: arcs. Animations for all thirteen fluids are provided as supplementary material.}
\label{fig:patterns}
\end{figure}

The mixing cycle is the same for all fluids (Fig.~\ref{fig:patterns}). During each half period one arc captures a tongue of the stream on its side and draws it into the vortex it drives (Sec.~\ref{sec:flow}); when the rotation reverses, the tongue is cut off, stretched around the neighbouring arc and folded, and is released downstream as a laminated lobe which rolls up behind the last arc. Each period thus delivers two folded lobes, one of each composition. Shear thinning changes how deep and how thin the tongues become: for $N=0.4$ they remain thick and penetrate less far into the vortices, and a large, nearly unmixed region of lean fluid persists between the first lower arc and the upper arc during half of the period; for $N=1$ they are drawn out further, wrapped once more around the vortices and leave as thinner striations.

\subsection{Outlet statistics}\label{subsec:outlet}
\begin{figure}[htbp]
\centering
\includegraphics[width=\textwidth]{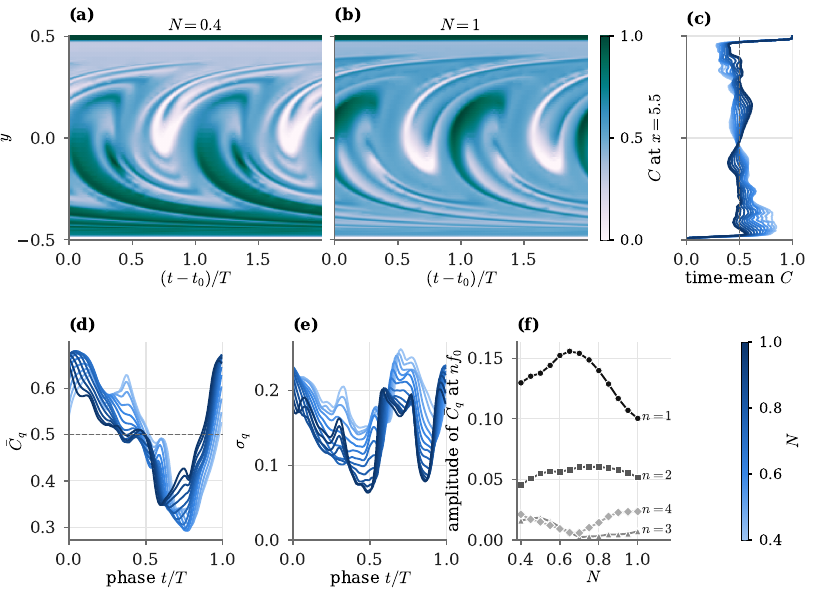}
\caption{The mixture delivered at the outlet $x=5.5$, periodic regime. (a, b) Space--time diagrams of the concentration over two forcing periods ($t_0=8$) for the lowest and the highest $N$. (c) Time-mean concentration profile. Phase-averaged (d) cup-mixing concentration $\bar C_q$ and (e) cross-sectional deviation $\sq$, Eq.~\eqref{eq:sigmaq}, over one period; colour: $N$. (f) Amplitude of the first four harmonics $nf_0$ of $\bar C_q(t)$ versus $N$.}
\label{fig:outlet}
\end{figure}

The mixture delivered at the outlet is periodic but not uniform. Each period delivers one crescent-shaped lobe of rich fluid and one of lean fluid, both laminated by thinner striations, with nearly steady streaks near each wall (Fig.~\ref{fig:outlet}a,b). The cup-mixing concentration $\bar C_q(t)$ therefore oscillates at the forcing frequency $f_0=1/T$, with a peak-to-peak amplitude of 0.30--0.38 (Fig.~\ref{fig:outlet}d), and more than 99.9\,\% of its variance is carried by the harmonics of $f_0$; $\sq(t)$ is modulated mainly at $2f_0$ (Fig.~\ref{fig:outlet}e). The time-mean profile (Fig.~\ref{fig:outlet}c) reveals a persistent segregation: in the outer quarters of the channel the mean concentration departs from $\tfrac12$ by up to 0.3, the fluid near each wall being on average of the composition opposite to its inlet one, a signature of the transverse transport by the arcs. The three to five cells adjacent to each wall carry nearly unmixed inlet fluid; this layer is set by the grid and carries less than 1\,\% of the flux. The amplitude of the fundamental of $\bar C_q$ is non-monotonic in $N$ (Fig.~\ref{fig:outlet}f): 0.13 for $N=0.4$, 0.16 at $N=0.65$ and 0.10 for $N=1$.

\begin{figure}[htbp]
\centering
\includegraphics[width=\textwidth]{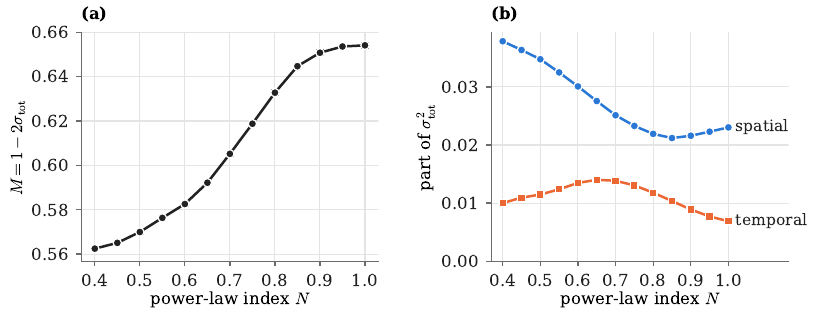}
\caption{Mixing at the outlet versus the power-law index $N$, periodic regime $5\le t\le10$. (a) Mixing efficiency $M=1-2\st$, Eq.~\eqref{eq:sigmatot}; the bars, smaller than the symbols, give the standard deviation of $M$ computed period by period over the ten periods. (b) Spatial and temporal parts of $\st^2$, Eq.~\eqref{eq:split}.}
\label{fig:sigmaN}
\end{figure}

\subsection{Mixing efficiency versus the power-law index}\label{subsec:MN}
Shear thinning degrades the mixing monotonically but not uniformly (Fig.~\ref{fig:sigmaN}a). $M$ decreases from 0.654 for the Newtonian fluid to 0.562 for $N=0.4$, i.e.\ by 14\,\%, while its period-to-period standard deviation is about $10^{-3}$, so that all differences between neighbouring $N$ are significant. The dependence is sigmoidal: between $N=1$ and 0.9 the efficiency hardly changes ($\Delta M=-3\times10^{-3}$), most of the loss occurs between $N=0.85$ and 0.6 ($\Delta M=-0.062$), and the curve flattens again for $N\le0.55$ ($\Delta M=-0.014$), where the wall viscosity approaches the plateau $\eta_\infty$ (Sec.~\ref{subsec:penetration}).

The two parts of the variance, Eq.~\eqref{eq:split}, respond differently (Fig.~\ref{fig:sigmaN}b). The spatial part decreases from $0.038$ for $N=0.4$ to a minimum of $0.021$ at $N=0.85$ and rises slightly for $N=1$ ($0.023$); the temporal part is largest at intermediate $N$ ($0.014$ for $N=0.65$) and smallest for $N=1$ ($0.007$). The improvement with $N$ thus has two successive origins: for $N\lesssim0.75$ a better cross-sectional homogenisation, which more than compensates a growing pulsation; for $N\gtrsim0.75$ the damping of the pulsation, as the stronger vortices also mix successive pulses with one another. Downstream of the reverse-flow region $\st$ decreases by 3.0--4.5\,\% between $x=4.75$ and the outlet, by the same relative amount for all $N$, so that the ranking of the fluids is set at the exit of the active zone (supplementary Sec.~S3).

\section{Effect of shear thinning on the flow}\label{sec:flow}

\subsection{Flow structure and flow type}
\begin{figure}[htbp]
\centering
\includegraphics[width=\textwidth]{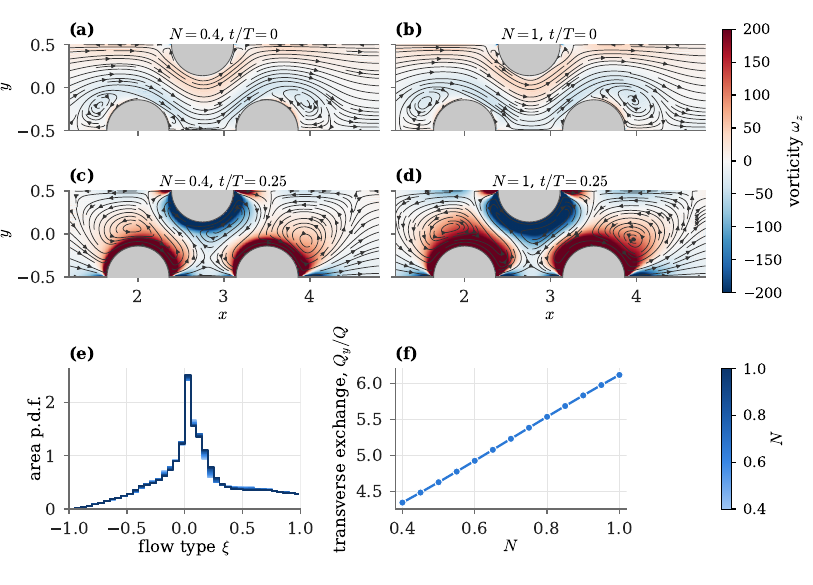}
\caption{Flow structure, last forcing period. (a--d) Vorticity and instantaneous streamlines with the arcs at rest ($t/T=0$) and at maximum rotation ($t/T=\tfrac14$) for the lowest and the highest $N$. (e) Area-weighted distribution of the flow-type parameter $\xi$, Eq.~\eqref{eq:flowtype}, over the fluid of the active zone $1.5<x<4.25$, 20 phases of the period; colour: $N$. (f) Transverse exchange across the centreline, relative to the axial flow rate.}
\label{fig:flow}
\end{figure}

With the arcs at rest ($t/T=0$; Fig.~\ref{fig:flow}a,b) the pressure-driven flow meanders around the arcs, leaving slowly recirculating cells in the cavities next to the lower arcs. At maximum rotation (Fig.~\ref{fig:flow}c,d) each arc drags a large vortex of its own sense of rotation, and the axial flow is squeezed between these vortices; half a period later the senses of rotation are reversed. This alternation is the same for all $N$; for $N=0.4$ the vorticity is more concentrated in the layers attached to the arcs and the vortices are weaker away from them.

The transverse exchange across the centreline over the active zone, $Q_y=\langle\int|v(x,0,t)|\dd x\rangle_t$, exceeds the axial flow rate four- to six-fold and increases linearly with $N$, from $Q_y/Q=4.34$ for $N=0.4$ to 6.12 for $N=1$ (Fig.~\ref{fig:flow}f): the more shear-thinning fluid is stirred less vigorously. The local flow type is, in contrast, insensitive to the rheology (Fig.~\ref{fig:flow}e): $\xi$ is sharply peaked at simple shear, with 17\,\% of the area in extension-dominated ($\xi>0.5$) and 5.5\,\% in rotation-dominated ($\xi<-0.5$) regions, and its mean varies only from 0.108 to 0.105 between $N=0.4$ and 1. Shear thinning thus changes the intensity of the stirring, not its kinematic nature; since stirring is governed by the rate of deformation rather than by the vorticity \cite{branicki2010}, it is this intensity that the rest of the analysis follows.

\subsection{Viscosity field and transmission of the wall forcing}\label{subsec:penetration}
\begin{figure}[htbp]
\centering
\includegraphics[width=\textwidth]{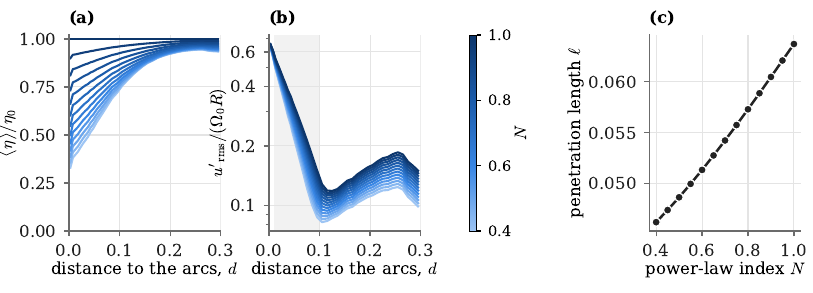}
\caption{Transmission of the wall forcing into the bulk, periodic regime, active zone $1.5<x<4.25$. (a) Time-mean viscosity and (b) amplitude of the oscillating velocity $u'_{\mathrm{rms}}/(\Omega_0R)$ versus the distance $d$ to the nearest arc surface; the shaded range in (b) is used to fit the e-folding length $\ell$. (c) Penetration length $\ell$ versus $N$.}
\label{fig:penetration}
\end{figure}

The arcs shear the adjacent fluid at rates of order $10^3$ (time-mean wall value $1.0$--$1.3\times10^3$), against 40--70 in the bulk, i.e.\ at local Carreau numbers $\lambda\dot\gamma\simeq10$--13 against about 0.5. The viscosity is therefore strongly reduced in a layer covering the arcs (Fig.~\ref{fig:penetration}a): its time mean at the wall falls from 0.90 for $N=0.95$ to 0.33 for $N=0.4$ (instantaneous minimum 0.15), while the bulk value stays above $0.94\,\eta_0$. The thickness of this layer, at which $\langle\eta\rangle$ recovers half of its drop, is nearly independent of $N$ ($\delta_\eta=0.062$--0.072). The oscillating velocity decays across it (Fig.~\ref{fig:penetration}b), from $0.53$--$0.58\,\Omega_0R$ at $d=0.01$ to $0.08$--$0.14\,\Omega_0R$ at $d=0.1$, nearly exponentially, with an e-folding length $\ell$ that increases monotonically from 0.046 for $N=0.4$ to 0.064 for $N=1$ (Fig.~\ref{fig:penetration}c): the more shear-thinning the fluid, the more closely the forcing is confined to the arcs.

\paragraph{A geometric length shortened by the wall viscosity} To identify what sets $\ell$, the fluids with $N=1$, 0.6 and 0.4 were also simulated at seven Strouhal numbers, $\St=0.25$--4 ($T=2$--0.125, $\Omega_0=150$), and with the rotation rate halved ($\Omega_0=75$) and doubled ($\Omega_0=300$) at $\St=1$ (Sec.~\ref{subsec:forcing}). Two classical candidates are ruled out. The Stokes layer of the oscillating wall, $(2\nu_0/\omega)^{1/2}\simeq0.40$, is six to nine times larger than $\ell$, and doubling the period leaves the Newtonian $\ell$ unchanged (0.0639 against 0.0637) whereas a Stokes layer would thicken by 41\,\%; built on the wall viscosity of $N=0.4$ it is still five times larger than $\ell$. A rotation layer $(\nu/\Omega_0)^{1/2}$ is ruled out likewise: halving $\Omega_0$ changes the Newtonian $\ell$ by less than 1\,\% (0.0642) and doubling it by $-2.6$\,\% (0.0621), against $+41$ and $-29$\,\% for such a layer. Over $\Rey_w=27$--108 and $\St=0.25$--2 the Newtonian penetration length is thus a property of the geometry of the forced flow, $\ell_0\simeq0.064\simeq0.18R$, as for a creeping flow driven by a moving boundary; the small decrease at the fastest forcings is the only trace of inertia.

\begin{figure}[htbp]
\centering
\includegraphics[width=\textwidth]{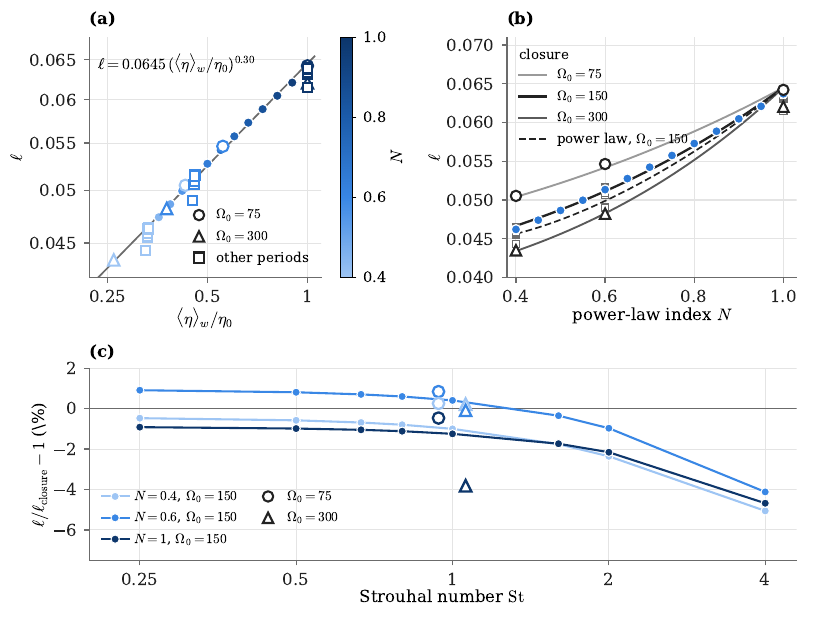}
\caption{Scaling of the penetration length. (a) $\ell$ versus the time-mean wall viscosity (logarithmic axes) for all 40 runs; filled circles: reference forcing $\Omega_0=150$, $\St=1$ (colour: $N$); open circles and triangles: $\Omega_0=75$ and 300 at $\St=1$; open squares: $\Omega_0=150$ at the other Strouhal numbers, $\St=0.25$--4; line: Eq.~\eqref{eq:ell_collapse}. (b) $\ell$ versus $N$: Eqs.~\eqref{eq:ell_collapse}--\eqref{eq:ell_closure} at $\Omega_0=75$, 150 and 300 (solid lines), power-law limit at $\Omega_0=150$, Eq.~\eqref{eq:ell_powerlaw} (dashed), and simulations (symbols as in (a); small grey squares: other Strouhal numbers). (c) Relative deviation of the simulated $\ell$ from Eqs.~\eqref{eq:ell_collapse}--\eqref{eq:ell_closure} versus $\St$ for $N=0.4$, 0.6 and 1 at $\Omega_0=150$ (lines), and at $\Omega_0=75$ and 300 (open symbols, shifted slightly about $\St=1$ for legibility). Constants $\ell_0=0.0645$, $\beta=0.295$, $k=41.3$, fitted on the shear-thinning fluids at the reference forcing.}
\label{fig:scaling}
\end{figure}

Shear thinning shortens this length through the viscosity drop next to the arcs. Over all 40 runs --- the thirteen fluids at the reference forcing and the runs at other rotation rates and periods --- $\ell$ is a function of the time-mean wall viscosity alone,
\begin{equation}\label{eq:ell_collapse}
\ell=\ell_0\left(\frac{\langle\eta\rangle_w}{\eta_0}\right)^{\beta},\qquad \ell_0=0.0645,\quad \beta=0.295,
\end{equation}
within 2\,\% for the shear-thinning fluids and 4\,\% for the Newtonian one up to $\St=2$ (Fig.~\ref{fig:scaling}a), $\ell_0$ and $\beta$ being fitted on the twelve shear-thinning fluids at the reference forcing only. Other measures of the viscosity of the layer do not collapse the runs at different forcings: the viscosity at the edge of the layer, $\eta(\ell)$, collapses the thirteen fluids as $\ell\propto\eta(\ell)^{1/2}$, but with an exponent that changes with $\Omega_0$ (0.64 at $\Omega_0=75$, about 0.42 at 300). The reach of the forcing is controlled by the viscosity where the momentum is injected: the thinned fluid acts as a lubricating film which partially decouples the bulk from the moving walls. The exponent $\beta$ is empirical, and depends on the definition of $\ell$: for six alternative definitions (other fit ranges, or the distance at which $u'_{\mathrm{rms}}$ falls by $e$ or 2) the same law holds within 2.2--6\,\%, with $\beta=0.19$--0.38 (supplementary Sec.~S4).

\paragraph{Wall closure} The law becomes predictive once the wall viscosity is expressed through the forcing. The wall shear rate follows the wall speed, $\dot\gamma_w(t)=k\,\Omega_0R\,|\sin\omega t|$; since the arcs stop twice per period, the wall viscosity recovers periodically towards $\eta_0$, and its time mean is the cycle average of the Carreau law,
\begin{equation}\label{eq:ell_closure}
\langle\eta\rangle_w=\frac{1}{\pi}\int_0^{\pi}\eta\big(k\,\Omega_0R\,\sin\psi\big)\,\dd\psi ,\qquad k=41.3 .
\end{equation}
Fitted on the twelve shear-thinning fluids at the reference forcing, Eq.~\eqref{eq:ell_closure} reproduces their wall viscosity within 1.3\,\%, and that of the other runs within 1.5\,\% up to $\St=2$. Together, Eqs.~\eqref{eq:ell_collapse}--\eqref{eq:ell_closure} give $\ell$ explicitly in terms of the rheology and the forcing, with three constants: within 1.3\,\% for the thirteen fluids (Fig.~\ref{fig:scaling}b), and within 2.4\,\% (shear-thinning) and 4\,\% (Newtonian) for the other forcings up to $\St=2$ (Fig.~\ref{fig:scaling}c). The value $1/k=0.024$ spans about four finest cells, so that $k$ partly reflects the discretisation (the wall viscosity changes by 6\,\% between levels 9 and 10, which changes $\ell$ by less than 2\,\%). The closure shear rate, of time mean $2k\Omega_0R/\pi=1.4\times10^3$, is thus an effective rate fitted on the viscosity rather than the wall shear rate itself, which is 10--40\,\% lower and varies with $N$. Evaluating the viscosity at the time-mean wall shear rate instead underestimates $\langle\eta\rangle_w$ by 15\,\% for $N=0.4$: the layer ``breathes'', recovering at each reversal.

\paragraph{Power-law limit} For $\lambda\dot\gamma\gg1$ and $\eta_\infty\to0$, Eqs.~\eqref{eq:ell_collapse}--\eqref{eq:ell_closure} take the closed form
\begin{equation}\label{eq:ell_powerlaw}
\frac{\ell}{\ell_0}=\Big[K(N)\,\Cu_w^{\,N-1}\Big]^{\beta},\qquad
\Cu_w=\lambda\,k\,\Omega_0R,
\end{equation}
where $\Cu_w\simeq22$ and $K(N)=\Gamma(N/2)/\big[\sqrt{\pi}\,\Gamma\big((N+1)/2\big)\big]$ is the cycle average of $|\sin\psi|^{N-1}$. The penetration length then depends on the rotation rate as $\ell\propto\Omega_0^{\beta(N-1)}$: not at all for a Newtonian fluid and weakly for shear-thinning ones ($\Omega_0^{-0.12}$ for $N=0.6$, $\Omega_0^{-0.18}$ for $N=0.4$; the runs give $\Omega_0^{-0.09}$ and $\Omega_0^{-0.11}$). Equation~\eqref{eq:ell_powerlaw} is within 3\,\% of the simulations at the reference forcing and 7\,\% at $\Omega_0=300$ (Fig.~\ref{fig:scaling}b), the departures coming from the bounds $\eta_0$ and $\eta_\infty$ of the Carreau law.

\subsection{Dependence on the forcing}\label{subsec:forcing}
The period does not change the wall viscosity, which is set by the wall speed, nor the penetration length of the shear-thinning fluids (0.0506--0.0516 for $N=0.6$ over $\St=0.25$--2), apart from a slow systematic decrease of 1.3--1.9\,\% between $\St=0.25$ and 2. At $\St=4$, however, $\ell$ is 3.5--4.5\,\% shorter than at $\St=1$ for all three fluids, as the near-wall flow no longer adjusts to the wall speed within a period; Eqs.~\eqref{eq:ell_collapse}--\eqref{eq:ell_closure}, which contain no frequency, should not be used beyond $\St\approx2$. The rotation rate acts through the thinning it produces: halving it raises the wall viscosity from 0.46 to 0.56 for $N=0.6$ and from 0.33 to 0.43 for $N=0.4$, lengthening $\ell$ by 6 and 9\,\%; doubling it lowers the wall viscosity to 0.38 and 0.26 and shortens $\ell$ by 6\,\%.

\begin{figure}[htbp]
\centering
\includegraphics[width=\textwidth]{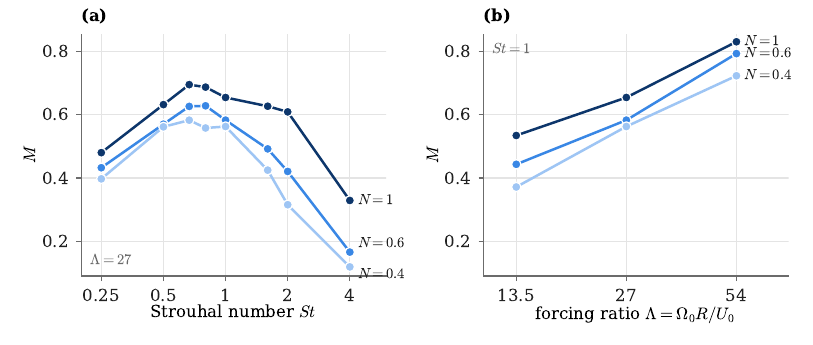}
\caption{Outlet mixing efficiency versus the forcing, for $N=1$, 0.6 and 0.4. (a) Strouhal number, $\Omega_0=150$ ($\Lambda=27$); (b) forcing ratio $\Lambda=\Omega_0R/U_0$, $\St=1$.}
\label{fig:forcing}
\end{figure}

\paragraph{Amplitude} The mixing efficiency depends on the forcing at fixed $\ell$ (Fig.~\ref{fig:forcing}b). Halving the rotation rate lowers $M$ from 0.654 to 0.534 for $N=1$, although $\ell$ is unchanged, from 0.583 to 0.443 for $N=0.6$ and from 0.562 to 0.372 for $N=0.4$; doubling it raises $M$ to 0.830, 0.793 and 0.723, the last above the Newtonian value at the reference forcing although its penetration is shorter. The penalty of shear thinning decreases as the forcing increases: $M(0.4)/M(1)=0.70$, 0.86 and 0.87, and $M(0.6)/M(1)=0.83$, 0.89 and 0.96, at $\Omega_0=75$, 150 and 300. The penetration length therefore predicts the effect of the rheology at a given forcing, whereas the level of mixing is set by the wall speed and by the period.

\paragraph{Period} The dependence on the period is non-monotonic (Fig.~\ref{fig:forcing}a). For $N=1$, $M=0.480$, 0.632, 0.695, 0.687, 0.654, 0.626, 0.609 and 0.329 at $\St=0.25$, 0.5, 0.67, 0.8, 1, 1.6, 2 and 4: a sharp maximum close to $\St=0.7$ and a collapse at fast forcing. For $N=0.6$ the largest values are at the same periods ($M=0.626$ and 0.628 at $\St=0.67$ and 0.8), but up to $\St=1$ shear thinning lowers $M$ by a nearly constant 9--11\,\%, whereas above the optimum the penalty grows to 21, 31 and 50\,\% at $\St=1.6$, 2 and 4. For $N=0.4$ ($M=0.397$, 0.561, 0.582, 0.558, 0.562, 0.425, 0.316 and 0.119) the maximum flattens into a plateau over $0.5\le\St\le1$, again highest at $\St=0.67$, and the penalty grows from 11--19\,\% to 32, 48 and 64\,\% at $\St=1.6$, 2 and 4. Shear thinning thus narrows the range of efficient periods from the side of fast forcing, without moving the optimum.

Two kinematic conditions bound this range. First, a parcel that crosses the mixing zone in less than half a period is displaced without being folded; every parcel is reversed at least once if $\St\ge\St_{\min}=W/(2U_0\tau_{\min})$, where $\tau_{\min}$ is the shortest residence time (Sec.~\ref{sec:lagrangian}). Second, the fluid at mid-channel must be displaced by about one channel width per half period; if the centreline transverse velocity oscillates with amplitude $\alpha\Omega_0R$, this requires $\St\lesssim\St_{\max}=\alpha\Lambda/\pi$. The measured fraction $\alpha$ is 0.283, 0.216 and 0.185 for $N=1$, 0.6 and 0.4, independent of the period up to $\St=2$, but it decreases with the rotation rate, the more so the stronger the shear thinning (for $N=0.4$, 0.214, 0.185 and 0.135 at $\Omega_0=75$, 150 and 300; for $N=1$, 0.291, 0.283 and 0.262): it is the counterpart, in the core of the channel, of the shortening of $\ell$. With $\tau_{\min}=0.40$, 0.48 and 0.52, the window $\St_{\min}\lesssim\St\lesssim\St_{\max}$ is $0.62$--2.4, 0.53--1.9 and 0.48--1.6 for $N=1$, 0.6 and 0.4: shear thinning lowers its upper limit much more than its lower one. At $\St=2$ the Newtonian fluid is inside the window, $N=0.6$ just outside and $N=0.4$ well outside it, the order of the losses observed. The Newtonian optimum lies near the lower limit, the slowest forcing that reverses every parcel (the fraction of the flux crossing within half a period is 0.44, 0.08 and 0 at $\St=0.25$, 0.5 and 0.8). Since a larger $\Lambda$ raises $\St_{\max}$, a stronger forcing moves the efficient range towards faster forcing, as observed in the experiments with shear-thinning solutions \cite[Sec.~5.2]{kiwan2025}. Those experiments agree with the simulations in their main features --- the fluids mixed alike under slow forcing, the solutions fell behind above an intermediate $\St$, in the order of their power-law index, and the Newtonian optimum was at $\St\approx0.7$ --- but placed the optimum of the shear-thinning solutions at $\St\approx0.5$, a shift that the simulations do not reproduce; the arc geometry, the forcing ratio (10), the Reynolds number ($\approx0.1$) and the finite depth of the experimental channel differ.

\subsection{Energy budget}
\begin{figure}[htbp]
\centering
\includegraphics[width=\textwidth]{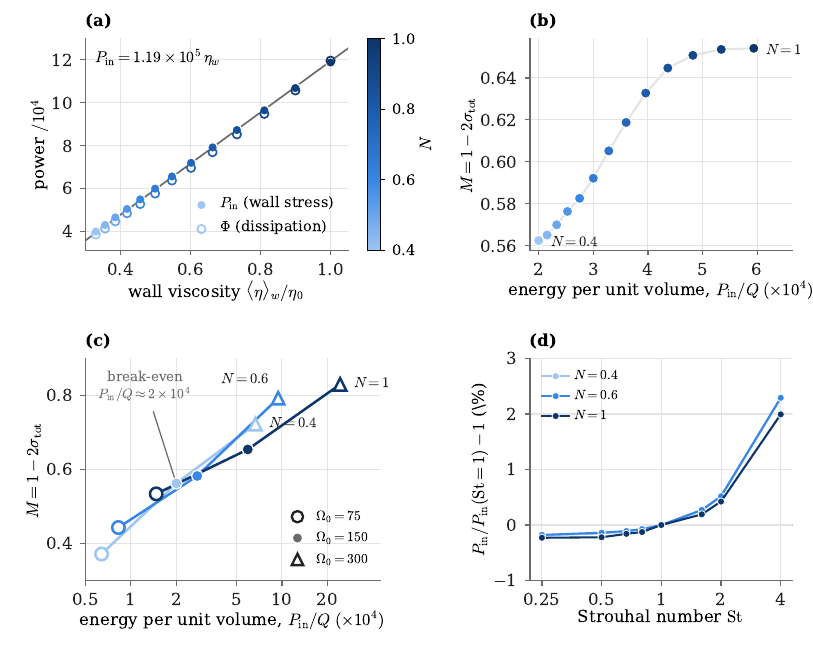}
\caption{Energy cost of the mixing, periodic regime. (a) Power injected by the arcs, computed from the wall stress (filled symbols), and viscous dissipation over the channel (open symbols), versus the time-mean viscosity at the arc surfaces; line: Eq.~\eqref{eq:power}. (b) Mixing efficiency at the outlet versus the energy spent per unit volume of fluid processed, reference forcing. (c) The same for $N=0.4$, 0.6 and 1 at $\Omega_0=75$, 150 and 300, $\St=1$ (logarithmic axis). (d) Injected power relative to its value at $\St=1$ versus the Strouhal number, $\Omega_0=150$. Colour: power-law index $N$.}
\label{fig:energy}
\end{figure}

The power injected by the arcs falls with shear thinning, from $P_{\mathrm{in}}=1.19\times10^5$ for $N=1$ to $4.0\times10^4$ for $N=0.4$ (Fig.~\ref{fig:energy}a). The dissipation accounts for 96--101\,\% of it, the deficit being attributable to the feet of the arcs, where the stress is singular and through which 24\,\% ($N=0.4$) to 41\,\% ($N=1$) of the power is injected. The power is controlled by the wall viscosity:
\begin{equation}\label{eq:power}
P_{\mathrm{in}}=C_P\,\langle\eta\rangle_w,\qquad C_P=1.19\times10^5,
\end{equation}
within 2\,\% over the thirteen fluids (a free fit gives the exponent 0.98): the wall kinematics is essentially fixed by the imposed rotation, the mean wall shear rate varying by 26\,\% while the wall viscosity varies threefold. Over all the forcing runs, the dissipation in the active zone equals $38.3\,\langle\eta\rangle_w(\Omega_0R)^2$ within 3\,\%.

The cost and the reach of the forcing are thus set by the same quantity with very different exponents, $P_{\mathrm{in}}\propto\langle\eta\rangle_w$ and $\ell\propto\langle\eta\rangle_w^{0.3}$. From $N=1$ to 0.4 the power decreases by 66\,\%, $\ell$ by 27\,\% and $M$ by only 14\,\%: per unit volume processed, the more shear-thinning fluids are mixed less well but much more cheaply (Fig.~\ref{fig:energy}b), the efficiency per unit energy $M\,Q/P_{\mathrm{in}}$ rising from $1.10\times10^{-5}$ to $2.81\times10^{-5}$, a factor 2.6. Conversely, between $N=0.9$ and 1, $P_{\mathrm{in}}/Q$ increases by 23\,\% for a gain of $3\times10^{-3}$ in $M$.

The two forcing parameters act very differently on this balance. The period costs nothing: over $\St=0.25$--4 the power changes by less than 2.5\,\% (Fig.~\ref{fig:energy}d), while $M$ varies by a factor two or more. The rotation rate is expensive (Fig.~\ref{fig:energy}c): doubling it from 150 to 300 multiplies the energy per unit volume by 4.1 for $N=1$ and by 3.4 and 3.3 for $N=0.6$ and 0.4, whose wall viscosity falls further, for gains of 0.18, 0.21 and 0.16 in $M$. Compared at equal energy rather than at equal forcing (interpolating logarithmically between the rotation rates), the fluid with $N=0.4$ reaches the Newtonian efficiency at $P_{\mathrm{in}}/Q\approx2\times10^4$ ($M\approx0.56$) and exceeds it above: at the energy of the Newtonian reference run, $5.9\times10^4$, it would reach $M\approx0.71$ against 0.654, as would $N=0.6$, which breaks even at about $3\times10^4$. Below $2\times10^4$ the Newtonian fluid remains ahead. The penalty of shear thinning is thus a penalty at a given rotation rate, which a faster rotation recovers at no extra energetic cost at moderate and large energy inputs. This comparison rests on three rotation rates and on the grid-limited $M$; since the relative penalty grows with $\Pe$ (Sec.~\ref{sec:lagrangian}), the break-even energy is expected to increase at larger $\Pe$.

\section{Lagrangian transport}\label{sec:lagrangian}

\subsection{Residence times}
\begin{figure}[htbp]
\centering
\includegraphics[width=\textwidth]{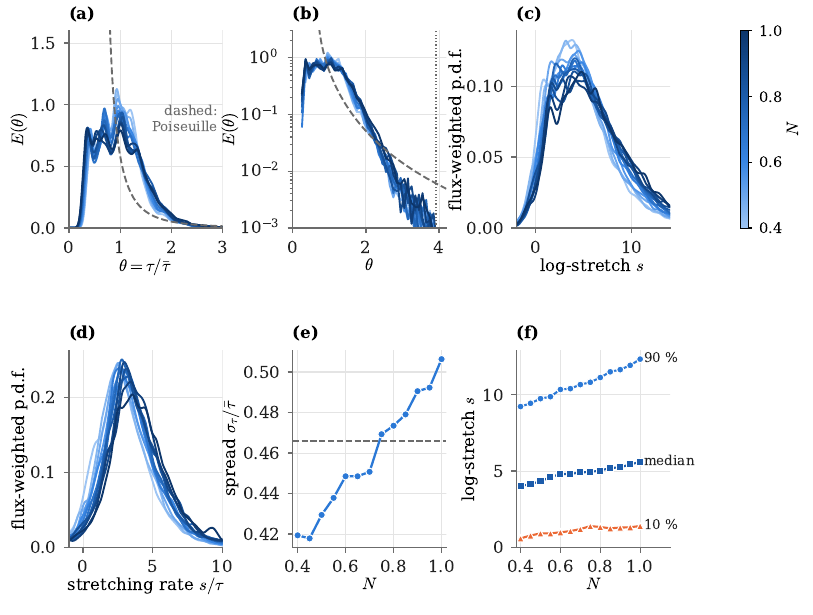}
\caption{Residence times and stretching of the fluid crossing the mixer between $x=1$ and $x=4.75$, periodic regime; flux-weighted distributions, colour: $N$. (a) Residence-time density $E(\theta)$, $\theta=\tau/\bar\tau$, and (b) the same in semi-logarithmic axes; dashed: plane Poiseuille flow, Eq.~\eqref{eq:rtd_pois}; dotted: end of the integration. (c) Density of the log-stretch $s$, Eq.~\eqref{eq:logstretch}, and (d) of the mean stretching rate $s/\tau$. (e) Spread of the residence time versus $N$; dashed: Poiseuille flow, over the same integration time. (f) 10th, 50th and 90th percentiles of $s$ versus $N$.}
\label{fig:rtd}
\end{figure}

Within twelve periods 99.71--99.88\,\% of the flux released at $x_{\mathrm{in}}$ has crossed $x_{\mathrm{out}}$, and the mean residence time, $\bar\tau=1.51$--1.55 for all $N$, agrees within 3.5\,\% with $V/Q=1.57$. The distributions (Fig.~\ref{fig:rtd}a,b) differ from that of plane Poiseuille flow, Eq.~\eqref{eq:rtd_pois}, in three respects. The fastest fluid, carried by the arcs, crosses in about a third of the mean time ($\theta_{\min}\simeq0.26$--0.34, against $\tfrac23$); the bulk of the distribution is modulated by the forcing, with maxima about one period apart, as the fluid is released in pulses; and the tail is exponential rather than algebraic, $E\propto e^{-\kappa_\tau\tau}$ with $\kappa_\tau=1.7$--2.0, the signature of escape from a chaotic region rather than of slow transit along the walls. The spread $\sigma_\tau/\bar\tau=0.42$--0.51 is a converged value, since less than 0.3\,\% of the flux remains in the channel after twelve periods. For Poiseuille flow the $\theta^{-3}$ tail makes the spread diverge logarithmically; evaluated in the same way, over the fluid that has left by the end of the integration (all but 1.2\,\% of the flux), it is 0.47, within the range of the simulations. The cross-sectional stirring of in-line mixers removes the slow wall fluid that dominates the residence times of a channel flow \cite{wiggins2004,PhysRevE.106.015107}, and replaces the algebraic tail by an exponential one.

Shear thinning leaves $\bar\tau$ unchanged but narrows the distribution: $\sigma_\tau/\bar\tau$ increases from 0.42 for $N=0.4$ to 0.51 for $N=1$ (Fig.~\ref{fig:rtd}e), and $\theta_{\min}$ rises from 0.26 for $N=1$ to 0.34 for $N=0.4$, consistent with a weaker exchange between the arcs and the core of the channel: less fluid is both short-circuited along the arcs and held back in their vortices. The experiments reported longer mean residence times for the shear-thinning solutions \cite[Sec.~5.2]{kiwan2025}, but as averages over tracers, which weight the slow fluid near the walls more heavily than the flux-weighted mean, which is fixed by mass conservation.

\subsection{Stretching}
The fluid is stretched very unevenly (Fig.~\ref{fig:rtd}c): the distribution of $s$ is broad and skewed, with a tail of strongly stretched fluid ($s>10$ for 8\,\% of the flux for $N=0.4$ and 18\,\% for $N=1$) and 15--25\,\% of the flux leaving with $s<2$. At most 0.3\,\% of the flux, trapped next to the arcs, reaches $s>30$; since this fraction is sensitive to the treatment of tracers at the walls, the distributions are characterised by quantiles. Shear thinning shifts the whole distribution towards weaker stretching (Fig.~\ref{fig:rtd}f): from $N=1$ to 0.4 the median of $s$ decreases from 5.6 to 4.0, its 10th percentile from 1.4 to 0.6 and its 90th percentile from 12.3 to 9.2. Since $\bar\tau$ does not change, this is due to a lower stretching rate, whose median falls from 3.9 to 2.7 (Fig.~\ref{fig:rtd}d), as in the closed eccentric-cylinder flow, where shear thinning increased the time constant of the stretching by up to 15\,\% \cite{ottino1}.

\begin{figure}[htbp]
\centering
\includegraphics[width=\textwidth]{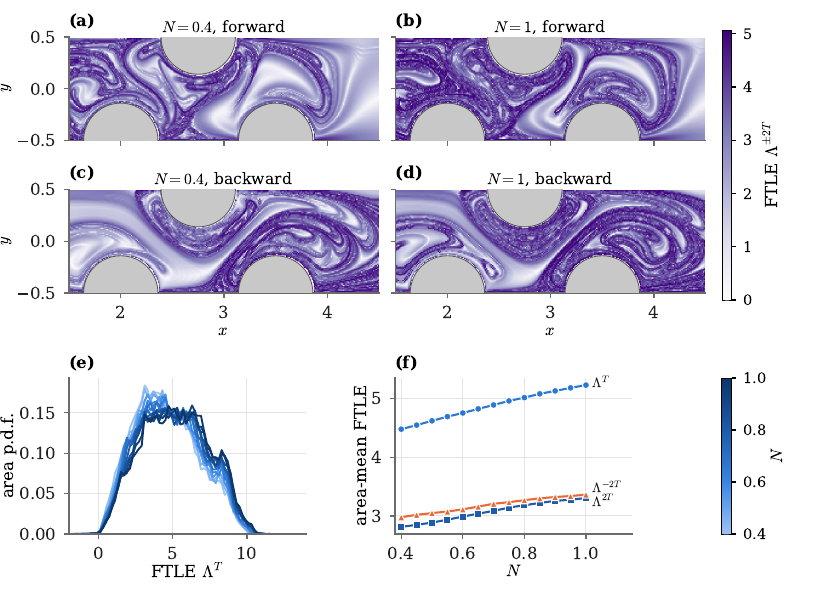}
\caption{Finite-time Lyapunov exponents, Eq.~\eqref{eq:ftle}, periodic regime. (a, b) Forward and (c, d) backward FTLE over two periods from phase $t/T=0$ for $N=0.4$ and $N=1$; grey: arcs. (e) Area distribution of the forward one-period FTLE over the active zone, two start phases; colour: $N$. (f) Area means of the one-period and of the forward and backward two-period FTLE versus $N$.}
\label{fig:ftle}
\end{figure}

\subsection{Finite-time Lyapunov exponents}
The FTLE fields (Fig.~\ref{fig:ftle}a--d) display the same ridges for the two extreme fluids: the forward ridges wrap around the arcs and fold in the cavities between them, and the backward ridges form the lobes ejected downstream of the second lower arc, which correspond to the folded filaments of the concentration field. Shear thinning does not move these structures but weakens them, as in the FTLE fields measured in this mixer \cite{kiwan2025}: for $N=0.4$ the regions of weak stretching, in particular the core of the passage between the arcs, are larger. The one-period FTLE is broadly distributed between 0 and 10 (Fig.~\ref{fig:ftle}e) --- finite-time exponents fluctuate widely, so that the distribution, not only its mean, characterises the stretching \cite{johnson2015} --- and its mean increases linearly from 4.5 to 5.2 between $N=0.4$ and 1 (Fig.~\ref{fig:ftle}f). The two-period exponents (2.8--3.4) follow the same trend, with forward and backward means within 6\,\% of each other, as in the experiments \cite[Sec.~5.2]{kiwan2025}.

\begin{figure}[htbp]
\centering
\includegraphics[width=\textwidth]{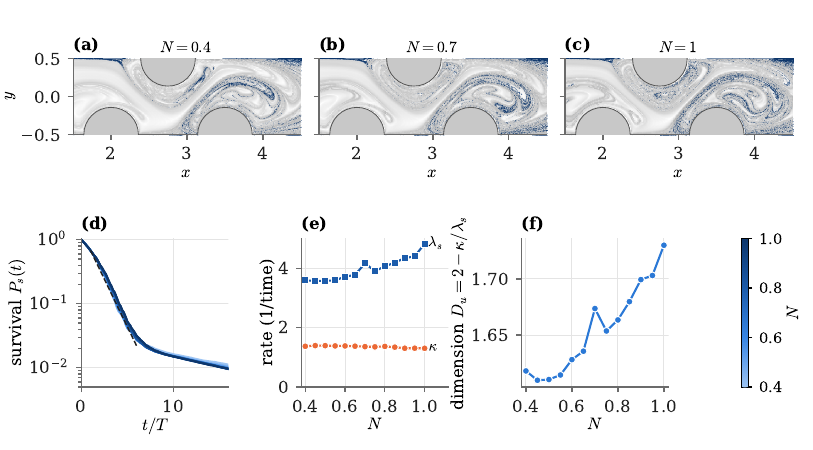}
\caption{Chaotic transport in the open flow. (a--c) Stroboscopic sections: positions at $t=nT$, $n=4,\dots,8$, of the tracers seeded uniformly over $1\le x\le4.75$ at $t=0$ and still in the mixing zone (blue), over the backward two-period FTLE (grey), for $N=0.4$, 0.7 and 1. (d) Survival probability $P_s(t)$, colour: $N$; dashed: exponential fit for $N=0.4$. (e) Escape rate $\kappa$ (fit over $T\le t\le6T$) and Lyapunov exponent of the saddle $\lambda_s$ versus $N$. (f) Dimension of the unstable manifold, $D_u=2-\kappa/\lambda_s$.}
\label{fig:chaos}
\end{figure}

\subsection{Chaotic saddle}
After four periods the surviving tracers outline a folded, filamentary set that coincides with the backward FTLE ridges --- the unstable manifold of the chaotic saddle --- concentrated in the vortex region between the upper arc and the second lower arc and in the lobes released downstream of it (Fig.~\ref{fig:chaos}a--c); the sections become richer in folds as $N$ increases. No regular island of significant area is found: after sixteen periods only 0.03--0.1\,\% of the tracers remain more than 0.02 away from a wall. The survival probability (Fig.~\ref{fig:chaos}d) decays exponentially for $t\lesssim6T$, as for a hyperbolic saddle \cite{kantz1985,lt2011}, with an escape rate $\kappa\simeq1.3$--1.4 almost independent of $N$ (1.37 to 1.30, within the $\pm0.1$ sensitivity to the fit window). Afterwards the decay slows down: about 1\,\% of the tracers remain after sixteen periods, more than 90\,\% of them within 0.02 of a wall, the effect of the no-slip walls \cite{gouillart2007,gouillart2008} in cells whose dynamics is not resolved here. The Lyapunov exponent of the saddle increases with $N$ from $\lambda_s=3.6$ to 4.8 (Fig.~\ref{fig:chaos}e), mostly above $N\simeq0.55$, so that $D_u$ increases from 1.61--1.62 for $N\le0.55$ to 1.73 for $N=1$ (Fig.~\ref{fig:chaos}f), robustly to the fit window ($\pm0.015$): the more shear-thinning the fluid, the less space-filling the set that carries the stretched and folded filaments out of the mixing zone \cite{jung1993,tel2005}.

\subsection{Lagrangian stirring and outlet mixing}
Across the thirteen fluids the outlet mixing efficiency follows the Lagrangian measures of stirring closely: $r^2=0.97$ with the mean one-period FTLE, 0.90 with the median stretching rate, 0.89 with the median log-stretch and with $D_u$, and 0.84 with $\lambda_s$, but only 0.75 with the poorly stretched fraction ($s<2$): the outlet variance is controlled by the stretching of the bulk of the flux. The Eulerian measures do as well: $M$ increases almost linearly with $\ell$ and with $Q_y$ ($r^2=0.97$), the dependence flattening only for $N\ge0.9$. The penetration length, the transverse exchange and the mean FTLE are three measures of one mechanism, the intensity with which the arc forcing is transmitted to the fluid, whereas the residence time and the escape rate, set by the geometry and the flow rate, hardly depend on the rheology. The scalar, however, cannot follow the stretching indefinitely: a striation of initial width $W/2$ reaches the grid scale once $s>\ln(W/2h)\simeq4.4$, as 44\,\% of the flux does for $N=0.4$ and 62\,\% for $N=1$, and is then homogenised by numerical diffusion. The outlet variance thus measures the fraction of the flux not stretched down to the grid scale, and its dependence on $N$ reflects the shift of the distribution of $s$.

\paragraph{Larger P\'eclet numbers} This argument can be extended to the P\'eclet numbers of applications with the lamellar model of the companion study \cite{burghelea_nwt}. Each parcel is a striation of initial thickness $\delta_0$ stretched by $e^s$ at a constant rate over its residence time; diffusion reduces its contrast by $(1+4\Theta)^{-1/2}$, with $\Theta=\mathcal{D}\tau(e^{2s}-1)/(2s\delta_0^2)$ \cite{ranz1979}, and $M=1-\sum_iw_i(1+4\Theta_i)^{-1/2}$, $w_i$ being the flux fraction of parcel $i$. With $\delta_0=0.23W$ and $\Pe_g=2.1\times10^4$ taken from the Newtonian study, without any fit on the shear-thinning runs, the model applied to the stretching distributions of the thirteen fluids reproduces $M(N)$ within 0.052 (rms 0.024), and inverted fluid by fluid it gives effective P\'eclet numbers of $0.95$--$2.3\times10^4$, confirming that the numerical diffusivity depends little on the rheology. The model overestimates the effect of shear thinning, $M(0.4)/M(1)=0.78$ against 0.86, but its trend is unambiguous: the ratio falls to 0.64 at $\Pe=10^6$ ($M=0.457$ and 0.294) and to 0.46 at $\Pe=10^9$, the order of the experiments \cite{kiwan2025} ($M=0.207$ and 0.095). At large $\Pe$ only fluid stretched beyond $s\simeq\ln(\delta_0\Pe^{1/2})$ is mixed, and the fraction above a high threshold decreases proportionally more when the distribution of $s$ shifts towards weaker stretching. The relative penalty of shear thinning is therefore larger at the P\'eclet numbers of applications, about twice the present 14\,\% at $\Pe\sim10^9$ even allowing for the bias of the model, so that this loss is a lower bound.

\section{Discussion}\label{sec:discussion}

\paragraph{Bulk kinematics} Shear thinning is not a qualitative perturbation of the stirring flow. Away from the arcs the viscosity stays above $0.94\,\eta_0$; the vortices, the Lagrangian ridges and the flow-type statistics are the same for all fluids, and the mean residence time and the escape rate are independent of $N$. The contrast with the closed eccentric-cylinder flow \cite{ottino1} is instructive: there the moving walls bound the whole domain, the mean viscosity falls to 0.3--0.8\,$\eta_0$, and velocity changes of a few per cent turn hyperbolic points into elliptic ones. Here the cross-flow at mid-channel weakens by a third between $N=1$ and 0.4 ($\alpha$ from 0.283 to 0.185), yet no island appears: in an open flow that the fluid crosses in a few periods, a weaker stirring shows up as a lower stretching rather than as new barriers to transport. The loss of the hyperbolic point between the arcs observed experimentally for the more shear-thinning solution under slow forcing \cite{kiwan2025} occurred at a forcing ratio of 10, where the cross-flow is weaker and presumably closer to the threshold of a topological change.

\paragraph{Confinement of the forcing} What shear thinning changes is the transmission of the arc rotation to the fluid. For a Newtonian fluid the forcing penetrates over a length $\ell_0$ set by the geometry, independent of the amplitude and of the period of the forcing; in a shear-thinning fluid the viscosity drop at the moving walls shortens it as $\ell=\ell_0(\langle\eta\rangle_w/\eta_0)^{0.3}$, the thinned layer partially decoupling the bulk from the walls. With the cycle-averaged wall viscosity, Eq.~\eqref{eq:ell_closure}, this law makes explicit the remark of Niederkorn and Ottino \cite{ottino1} that with a shear-thinning fluid the wall speed matters and not only the wall displacement \cite{swanson1990}: a faster rotation lowers the wall viscosity further and shortens the reach of the forcing. All the downstream consequences follow from this confinement: from $N=1$ to 0.4 the transverse exchange decreases by 29\,\%, the mean FTLE by 14\,\%, the Lyapunov exponent of the saddle by 26\,\%, $D_u$ from 1.73 to 1.61 and the median log-stretch from 5.6 to 4.0. This gives a quantitative form to the viscous slip invoked for the experiments \cite{kiwan2025}: the slip is the thinned layer at the arc surfaces, its effect is the shortening of $\ell$, and the fluid beyond the reach of the forcing is the fluid that remains poorly mixed.

\paragraph{Energetic cost and implications} The cost and the reach of the forcing are both set by the wall viscosity, the power in proportion and the reach through the power 0.3. Shear thinning therefore reduces the power much more than the mixing, and at equal energy per unit volume, with a faster rotation, a strongly shear-thinning fluid mixes as well as a Newtonian one, and better at larger energies: over the fourfold range of rotation rates simulated the energy grows by a factor 10.4 for $N=0.4$ and 11.4 for $N=0.6$, against 16.5 for $N=1$. For wall-driven mixers processing shear-thinning materials, the relevant quantity is thus neither the bulk viscosity nor the mean shear rate but the reach of the forcing. A faster rotation intensifies the stirring but does not extend its reach --- for $N=0.4$, $M$ rises from 0.372 to 0.723 when $\Omega_0$ is quadrupled while $\ell$ decreases as $\Omega_0^{-0.11}$ --- and the period, which costs nothing, must be chosen within a window that shear thinning narrows from the side of fast forcing. The part of the channel beyond the reach of the forcing can only be mixed better by bringing the moving walls closer to it. These consequences remain to be tested on other geometries.

\paragraph{Limitations} The simulations are two-dimensional; in a mixer of finite span the end walls add secondary flows and three-dimensional stretching. The rheology is purely viscous, without elasticity, thixotropy or yield stress, and the relaxation time and the viscosity ratio are fixed. The scalar is resolved down to a fixed grid scale, with an effective P\'eclet number of about $2\times10^4$, so that the conclusions rest on the variation of $M$ with $N$ and on Lagrangian measures independent of the scalar resolution; the lamellar model indicates that the relative penalty of shear thinning grows at larger $\Pe$. Finally, the fluid trapped next to the walls is not resolved.

\section{Conclusions}\label{sec:conclusions}
I have studied by two-dimensional simulations the mixing of Carreau fluids in the rotating arc-wall mixer for thirteen power-law indices between $N=1$ and 0.4, all other parameters being fixed.
\begin{enumerate}
\item Shear thinning degrades the mixing monotonically, from $M=0.654$ for $N=1$ to 0.562 for $N=0.4$, along a sigmoidal curve: a weak shear thinning ($N\ge0.9$) is harmless, and most of the loss occurs between $N=0.85$ and 0.6.
\item The bulk flow is hardly affected: the flow topology, the mean residence time and the escape rate from the chaotic saddle are independent of $N$.
\item Shear thinning acts through a thin layer around the moving arcs, where the viscosity falls by up to a factor three. The Newtonian penetration length of the forcing, $\ell_0\simeq0.18R$, is set by the geometry; shear thinning shortens it as $\ell=\ell_0(\langle\eta\rangle_w/\eta_0)^{0.3}$. With the wall viscosity given by the cycle-averaged Carreau law at a shear rate proportional to the wall speed, this law, fitted at one forcing, predicts $\ell$ within 4\,\% for rotation rates varied fourfold and periods eightfold ($\St=0.25$--2).
\item The confinement of the forcing reduces the transverse exchange, the finite-time Lyapunov exponents, the Lyapunov exponent and dimension of the chaotic saddle and the stretching; $\ell$, the transverse exchange and the mean FTLE each predict $M$ with $r^2=0.97$. A lamellar model predicts that the relative penalty of shear thinning grows with $\Pe$, to $M(0.4)/M(1)\approx0.5$ at $\Pe\sim10^9$.
\item The optimal Strouhal number, close to 0.7, does not move with $N$, but shear thinning narrows the range of efficient periods from the side of fast forcing, as explained by two kinematic conditions: every parcel reversed at least once, and a transverse displacement of one channel width per half period.
\item The injected power is proportional to the wall viscosity, so that the mixing efficiency per unit energy rises 2.6-fold from $N=1$ to 0.4; at equal energy per unit volume, with a faster rotation, the fluid with $N=0.4$ mixes as well as the Newtonian one at $P_{\mathrm{in}}/Q\approx2\times10^4$ and better above, whereas the period is a free control of the mixing.
\end{enumerate}
The penetration-length law provides a quantitative basis for the design of wall-driven mixers for shear-thinning materials; three-dimensional flows, elastic and yield-stress fluids and other wall-driven geometries are natural next steps.

\section*{CRediT authorship contribution statement}
\textbf{Teodor Burghelea:} Conceptualization, Methodology, Software, Formal analysis, Investigation, Visualization, Writing -- original draft, Writing -- review \& editing.

\section*{Declaration of competing interest}
The author has no known competing financial interests or personal relationships that could have appeared to influence the work reported in this paper.

\section*{Declaration of generative AI and AI-assisted technologies in the research and writing process}
During the preparation of this work the author used Claude (Anthropic), a large language model, in order to assist with running the simulations on the computing cluster, writing the analysis and plotting scripts, analysing the data, and drafting and editing the text. The author designed the study and checked the analyses and the results. After using this tool, the author reviewed and edited the content as needed and takes full responsibility for the content of the publication.

\section*{Data availability}
The data that support the findings of this study are available from the author upon reasonable request.

\section*{Supplementary material}
The supplementary material is provided as ancillary files of this arXiv submission. A supplementary document (\texttt{supplementary\_material.pdf}) describes the adapted grid and the numerical checks (S1), the establishment of the time-periodic regime (S2), the development of mixing along the channel (S3) and the sensitivity of the penetration-length law to the definition of $\ell$ (S4). Thirteen video files, \texttt{movie\_N0.40.mp4} to \texttt{movie\_N1.00.mp4}, show the concentration field over the ten forcing periods $5\le t<10$ of the periodic regime for each fluid (level-10 fields, 25 frames per second, one frame every $T/50$).

\section*{Acknowledgements}
I thank M.~Kiwan for discussions of the experiments in the rotating arc-wall mixer. This research used resources of the GLiCID Computing Facility (Ligerien Group for Intensive Distributed Computing, \url{https://doi.org/10.60487/glicid}, Pays de la Loire, France). This research did not receive any specific grant from funding agencies in the public, commercial, or not-for-profit sectors.

\bibliographystyle{elsarticle-num}
\bibliography{refs}

\end{document}